\documentclass[12pt]{article}

\pdfoutput=1
\usepackage{amsfonts,amsmath,amssymb}
\usepackage[paper=letterpaper,margin=1.0in]{geometry}
\usepackage{graphicx}
\usepackage{cite}
\usepackage{chngcntr}
\usepackage{tocloft}
\usepackage{hyperref}
\usepackage{xcolor}
\usepackage{mathtools}

\graphicspath{{figures/}}

\usepackage[sort,numbers]{natbib}

\newcommand\be{\begin{equation}}
\newcommand\ee{\end{equation}}
\newcommand\bea{\begin{eqnarray}}
\newcommand\eea{\end{eqnarray}}
\newcommand\ba{\begin{array}}
\newcommand\ea{\end{array}}
\newcommand\ben{\begin{enumerate}}
\newcommand\een{\end{enumerate}}
\newcommand\bi{\begin{itemize}}
\newcommand\ei{\end{itemize}}
\newcommand\bc{\begin{center}}
\newcommand\ec{\end{center}}
\newcommand\bfig{\begin{figure}}
\newcommand\efig{\end{figure}}
\newcommand\bq{\begin{quotation}}
\newcommand\eq{\end{quotation}}
\newcommand\bt{\begin{table}}
\newcommand\et{\end{table}}
\newcommand\btab{\begin{tabular}}
\newcommand\etab{\end{tabular}}

\newcommand\eref[1]{(\ref{#1})}

\newcommand\tr{\mathrm{Tr}}

\counterwithin*{equation}{section}
\renewcommand\thefootnote{\fnsymbol{footnote}}
\newcommand\comment[1]{}

\newcommand\rap{\eta}

\renewcommand\tilde{\widetilde}

\newcommand\sech{\mathrm{sech}}

\begin{document}

${}$

\vspace{.5cm}

\begin{center}
{\Huge \textbf{Entanglement, anomalies and Mathisson's helices}}

\vspace{.5cm}

Piermarco Fonda\footnote{\href{mailto:fonda@lorentz.leidenuniv.nl}{\texttt{fonda@lorentz.leidenuniv.nl}}}$^{1}$,
Diego Liska\footnote{\href{mailto:lis14392@uvg.edu.gt}{\texttt{lis14392@uvg.edu.gt}}}$^2$,
\'Alvaro V\'eliz-Osorio\footnote{\href{mailto:aveliz@gmail.com}{\texttt{aveliz@gmail.com}}}$^{3}$

\vspace{.33cm}

\vspace*{.2cm}
{${}^{1}$ Instituut-Lorentz, Universiteit Leiden,\\ P.O. Box 9506, 2300 RA Leiden, The Netherlands\\}

\vspace*{.2cm}
{${}^{2}$ Department of Physics, Universidad del Valle de Guatemala,\\
18 Avenida 11-95, Zona 15 Guatemala, Guatemala}

\vspace*{.2cm}
{${}^{3}$ M. Smoluchowski Institute of Physics, Jagiellonian University,\\
\L{}ojasiewicza 11, 30-348 Krak\'ow, Poland\\}

\end{center}

\vspace{1cm}

\centerline{\textbf{Abstract}}
We study the physical properties of a length-torsion functional which encodes the holographic entanglement entropy for 1+1 dimensional theories with chiral anomalies. Previously, we have shown that its extremal curves correspond to the mysterious Mathisson's helical motions for the centroids of spinning bodies. We explore the properties of these helices in domain-wall backgrounds using both analytic and numerical techniques. Using these insights we derive an entropic $c$-function $c_{\mathrm{Hel}}(\ell)$ which can be succinctly expressed in terms of  Noether charges conserved along these helical motions. 
 While for generic values of the anomaly
there is some ambiguity in the definition of $c_{\mathrm{Hel}}(\ell)$, we argue that at the chiral point this ambiguity is absent.
\bigskip

\newpage

\setcounter{tocdepth}{2}
\tableofcontents

\setcounter{footnote}{0}
\renewcommand\thefootnote{\arabic{footnote}}

\section{Introduction and summary}
\label{sec:intro}

Zamolodchikov's $c$-theorem \cite{Zamolodchikov:1986gt} provides a deep insight into the general properties of quantum field theories. It states that for every Poincar\'e invariant local theory in two dimensions there exists a function of the couplings that decreases monotonically along renormalization group (RG) flows. Furthermore, if the theory in question is conformal, this function is constant and matches the central charge of the conformal field theory (CFT). This theorem formalizes the intuition that information regarding the microscopic details of the theory is lost as one studies it macroscopically, meaning that RG flows are irreversible. It is possible to generalize this result to chiral theories \cite{Bastianelli:1996gh}: in this case, two distinct $c$-functions can be defined, one for left-movers and one for right-movers. As demonstrated in \cite{Bastianelli:1996gh}, the difference of these functions is constant along RG flows. This difference quantifies the number of degrees of freedom that are precluded from becoming massive at large distances. The fact that this quantity remains constant along the flow is a two-dimensional version of  't  Hooft's anomaly matching condition for the chiral anomaly. In contrast, the sum of these functions decreases monotonically along the RG just as the non-chiral $c$-function.

It is well-known that there is an intimate relationship between gravity in AdS$_3$ and two-dimensional CFTs. Indeed, as discovered by Brown and Henneaux \cite{Brown:1986nw}, the asymptotic symmetry algebra for Einstein gravity in AdS$_3$ corresponds to a left- and a right-moving Virasoro algebra with identical central charges. In this work, we are interested in theories with chiral anomalies for which there is a mismatch between left- and right- moving central charges. On the gravity side, this is implemented by regarding AdS$_3$ as a solution of Topologically Massive Gravity (TMG) \cite{Deser:1982vy}. In this case, the difference between the central charges is proportional to the Compton wavelength of the massive graviton \cite{Hotta:2008yq}. It has been argued that for arbitrary values of the graviton's mass, TMG violates either unitarity or positivity \cite{Li2008}. However, at the critical point where the graviton's Compton wavelength matches the AdS$_3$ radius these problems are absent. In this case, the dual central charges read 
\be\label{chira}
(c_L,c_R)=\left(0,\,3L/G_3\right)\,,
\ee
where $L$ is the AdS$_3$ radius and $G_3$ is the three-dimensional Newton's constant. Condition \eqref{chira} is known as the chiral point and the theory associated with it, called chiral gravity, has BTZ black holes and gravitons with non-negative mass \cite{Li2008}.

The physics of RG flows can be probed using quantum information theoretic quantities such as entanglement entropy (EE). For 1+1 dimensional quantum field theories, it is well-known that the EE of a space-like interval of length $\ell$ exhibits logarithmic divergences. Nonetheless, it is  possible to extract its universal part by computing the \emph{renormalized entanglement entropy} (REE) \cite{Casini:2004bw, Liu:2012eea}
\be
\widehat{S}_{EE}(\ell)= \ell \frac{d}{d\ell}S_{EE}(\ell)\,.
\label{REE1}
\ee
If the field theory in question is conformal then the REE is proportional to the central charge and for a RG flow interpolating between two CFTs  it was shown by Casini and Huerta \cite{Casini:2004bw} that \eqref{REE1} interpolates monotonically between their respective central charges, a result which furnishes an \emph{entropic} proof of Zamolodchikov's $c$-theorem. This story has a gravitational counterpart: it was shown by Ryu and Takayanagi (RT) \cite{Ryu:2006bv} that calculating entanglement entropy (EE) for field theories with a dual gravitational description involves finding a suitable extremal surface in an asymptotically AdS spacetime. Whenever the dual gravitational description corresponds to Einstein gravity, the relevant surface extremizes the area functional with properly chosen boundary conditions. In the case of 1+1 dimensional theories, the EE of a space-like interval of length $\ell$ is obtained through a bulk curve $\gamma(\ell)$ which extremizes the functional
\be  
{\cal F}_{0}[\gamma] =\mathfrak{m}\, \mathrm{L}[\gamma]\,,
\label{Len}
\ee
whose endpoints correspond to those of the boundary interval. The entanglement entropy associated to the interval is given by the length of this curve i.e. 
\be
S_{\mathrm{EE}}(\ell)={\cal F}_{0}[\gamma(\ell)]\,,
\label{RT}
\ee
for the appropriate value of $\mathfrak{m}$.
Using the gravitational duals of RG flows it is possible to construct an entropic $c$-function out of geodesic lengths \cite{Myers:2012ed}.

It is often the case that the dual gravitational description of a field theory does not correspond to Einstein gravity. Indeed, as mentioned above, 1+1 dimensional conformal field theories with chiral anomalies require a modification of the bulk theory, which calls for a generalization of the RT prescription. The authors of \cite{Castro:2014tta} have shown that, in this context, the holographic entanglement entropy is encoded by the functional 
\be  
{\cal F}[\gamma] =\mathfrak{m}\, \mathrm{L}[\gamma] + \mathfrak{s} \int_\gamma \tau\,,
\label{torsionaction1}
\ee
where $\tau$ is the extrinsic torsion of the curve $\gamma$, while $\mathfrak{m}$ and $\mathfrak{s}$ are suitably chosen constants. In the following,  we would like to proceed in a similar fashion as outlined in the previous paragraph.
First, we consider a boundary interval of proper length $\ell$ and join its endpoints $(X_\pm , T_\pm )$ with an extremum of \eqref{torsionaction1}. Then we must elucidate how the on-shell value of \eqref{torsionaction1} changes with $\ell$ in order to find the analogue of Eq.\,\eqref{REE1}. Difficulties arise from the fact that the extrema of \eqref{torsionaction1} correspond to Mathisson's helical motions of spinning bodies \cite{Fonda:2018eqg} which are, in general, more complicated than geodesics. Moreover, since Mathisson's helices are solutions to higher order equations, it is necessary to fix more boundary conditions in order to single out a solution. Nevertheless, these adversities can be surmounted and in the present work we show that this geometrical problem has a succinct solution. We find that the renormalized ${\cal F}[\gamma]$ is given by
\be
\widehat{\cal F}\equiv \ell \frac{d{\cal F}}{d\ell}= \left( T_+-T_-\right)\,\mathcal{Q}_t + \left( X_+-X_-\right) \,\mathcal{Q}_x\,,
\label{res1}
\ee
where $\mathcal{Q}_t $ and $\mathcal{Q}_x$ are conserved charges associated to rigid translations of relativistic spinning bodies. Equation \eqref{res1} is valid for any continuously varying family of Mathisson's helices connecting the interval's endpoints as we change $\ell$. In general, there exist an infinite number of such families. However, at the chiral point, a natural prescription emerges that allows to select a single family of helices to which we can associate the quantity
\be
\frac{c_{\mathrm{Hel}}(\ell)}{3} \equiv \widehat{\cal F}(\ell) =   \frac{\mathcal{Q}_x}{\mathcal{Q}_t} \,, 
\ee
which displays the features expected by an entropic \textit{c-}function for a wide variety of examples; namely, in a renormalization group flow  setting, it monotonically interpolates between two CFT central charges.

\section{Shape equations and anomalies}

In this Section we study the general properties of extremal curves associated with the functional \eqref{torsionaction1}. The following discussion requires some acquaintance with extrinsic geometric terminology, we refer the reader to \cite{Fonda:2016ine, Fonda:2018eqg} for a detailed discussion and notation. To describe the geometry of a curve embedded in a three-manifold we must introduce a \emph{moving frame} comprised by a normalized tangent vector $t^\mu$ and two normal vectors $n_A^{\;\mu}$, with $ A=1,2$, defined by 
\be\label{frame}
t_\mu\, n_A^{\;\mu}=0\,,\qquad  g_{\mu\nu}n_A^{\;\mu} n_B^{\;\nu}=\eta_{AB}\,,
\ee
where $\eta_{AB}=\mathrm{diag}(1,-1)$.
Using this moving frame we define the extrinsic curvatures and torsion:
\be\label{extrinsic curvature}
k^A=t^\mu\,{\cal D}_s\, n^{A}_\mu\,,\qquad
\tau=\frac{1}{2}\epsilon_{AB}\,\left( n^{A}_\mu\,{\cal D}_s\, n^{B\mu}\right)\,,
\ee
where ${\cal D}_s=t^\mu\nabla_\mu$ is the directional derivative along the curve.
In terms of these quantities, the equations that dictate the shape of the extremal curves can be written as \cite{Fonda:2018eqg}
\be
\mathfrak{m}\,k_A  
+
\mathfrak{s} \,  \epsilon_{AB}
\left[
(\tilde{D} k)^B
+
R_s^{\;B}\right] = 0\,,\label{shape torsion2}
\ee
where 
\be\label{gauge covariant}
(\tilde D V)^A=\partial_s V^A +\tau\, \epsilon^{AB}\eta_{BC} V^C\,,
\ee
and 
\be
R_s^{\;B}=t^\mu n^{\nu B}R_{\mu\nu}\,.
\ee
Using Frenet-Serret extrinsic quantities, which are those associated with a moving frame where the extrinsic curvature in one of the normal directions is set to vanish identically, the shape equations take the convenient form:
 \be\label{change in kFS}
\partial_s k^2_{\mathrm{FS}}=-2 R_s^{\;B}k_B \qquad  k^2_{\mathrm{FS}}\left(\mathfrak{m}-\mathfrak{s}\tau_{\mathrm{FS}}\right)=-\mathfrak{s}\epsilon_{AB}k^AR_s^{\;B}\,,
\ee
where $k^2_{\mathrm{FS}}=\eta_{AB}k^A k^B$ is the total curvature and $\tau_{\mathrm{FS}}$ is the torsion in the Frenet-Serret frame. From these equations it clearly follows that in a maximally symmetric ambient space, the total curvature is constant along extremal curves. Moreover, provided $k_{\mathrm{FS}}^2\neq 0$, the Frenet-Serret torsion reads
\be\label{tau}
\tau_{\mathrm{FS}}=\frac{\mathfrak{m}}{\mathfrak{s}}\,,
\ee
and it is thus fixed by the couplings of the theory.

It is important to point out that the \emph{shape equations} \eqref{shape torsion2} are closely related to the Mathisson-Papapetrou-Dixon (MPD) equations \cite{Mathisson:1937zz, Papapetrou:1951pa, Dixon:1970zz}
\begin{align}
&{\cal D}_s p^\lambda=-\frac{1}{2} t^\nu {\cal S}^{\rho\sigma} R^\lambda_{\;\;\nu\rho\sigma}\label{MPD1}\\
&{\cal D}_s {\cal S}^{\mu\nu}= p^\mu t^\nu-t^\mu p^\nu\,,\label{MPD2}
\end{align}
 for the momentum $p^\mu$ and spin ${\cal S}^{\mu\nu}$ of an extended body in the pole-dipole approximation.
The equivalence between the two systems holds only when the above are supplemented with the Mathisson-Pirani (MP) condition \cite{Mathisson:1937, Pirani:1956tn}
\be\label{MP}
{\cal S}^{\mu\nu}t_\mu=0\,, 
\ee
upon identifying 
\be
-p^\mu=\mathfrak{m}t^\mu+\mathfrak{s}\epsilon_{AB}k^An^{B\mu}\qquad  {\cal S}^{\mu\nu}=\mathfrak{s}\epsilon_{AB}n^{A\mu}n^{B\nu}\,.
\ee
Solutions $\gamma(s)$ to the system MPD with MP conditions are known in the literature as \emph{Mathisson's helices} \cite{Costa:2011zn}; hereafter we use this nomenclature for the extrema of \eqref{torsionaction1}.
Regarding the solutions to the shape equations \eqref{shape torsion2} as the trajectories of spinning bodies naturally evokes concepts from dynamics. For instance, the idea that symmetries of the ambient manifold induce conserved quantities along trajectories inexorably comes to mind. Concretely, to any Killing field $\xi_\mu$ in the ambient manifold we can associate a \emph{charge}
\be\label{conserved}
{\cal Q}_\xi[\gamma]=\xi_\mu p^\mu+\frac{1}{2}{\cal S}^{\mu\nu}\nabla_\mu \xi_\nu\,,
\ee
which is conserved along Mathisson helices \cite{Hojman:1976kn}. These charges will prove crucial in the arguments we develop in the forthcoming Sections. It is straightforward to verify that \eqref{conserved} is conserved using the MPD equations. Nevertheless, we provide a derivation of these charges using Noether's theorem in Appendix\,\ref{app:cha}\,.

\subsection{AdS$_3$  helices}
\label{sec:AdS hel}
The shape equations \eref{shape torsion2} can be solved analytically for maximally symmetric ambient manifolds. In \cite{Fonda:2018eqg} all possible solutions in AdS$_3$ spacetime,
\be
ds^2=\frac{L^2}{z^2}\left(\eta_{ab}\,dx^adx^b+dz^2\right)\,,
\ee
were classified in terms of their  Frenet-Serret curvature $k_{\mathrm{FS}}^2$ and torsion $\tau_{\mathrm{FS}}$. Moreover, since  the Frenet-Serret torsion is fixed uniquely by the couplings of the theory ($\tau_{\mathrm{FS}}=\mathfrak{m}/\mathfrak{s}$), the only free parameter is the curvature $k_{\mathrm{FS}}^2$.  For the purposes of the present discussion, we are concerned with solutions which connect space-like separated points in the boundary. This requirement imposes restrictions on the allowed values of $k_{\mathrm{FS}}^2$ relative to $\tau_{\mathrm{FS}}$ and the AdS$_3$ radius, either 
\be\label{Ia and c}
0 \leq k_{\mathrm{FS}}^2\leq \left(\frac{1}{L}-\tau_{\mathrm{FS}}\right)^2\qquad
\text{or }\qquad
 k_{\mathrm{FS}}^2\leq 0\,.
\ee
In the first case $|L \tau_{FS}| < 1$ and the solution must be isometric\footnote{Every curve with fixed  $ k_{\mathrm{FS}}^2$  and $ \tau_{\mathrm{FS}}$ in AdS$_3$ can be written as the image of an AdS$_3$  isometry acting on a seed solution, this follows from the fundamental theorem of curves. } to
\be\label{Ia}
\gamma^\mu_{I_a} =
\frac{L}{a \cosh \lambda_- s+b \cosh \lambda_+ s}
\left(
\begin{array}{c}
a \sinh \lambda_- s \\
b \sinh \lambda_+ s \\
L
\end{array} 
\right) \,,
\ee
where $a^2 + L^2 = b^2$ and $a^2 \lambda_-^2+1= b^2 \lambda_+^2$. In the latter instance, solutions are isometric to
\be
\gamma^\mu_{I_c} =
\frac{L}{a \sinh \lambda_- s+b \cosh \lambda_+ s}
\left(
\begin{array}{c}
a \cosh \lambda_- s \\
b \sinh \lambda_+ s \\
L
\end{array} 
\right) \,,
\label{Ic}
\ee
with $a^2 + b^2 = L^2$ and $a^2 \lambda_-^2 + b^2 \lambda_+^2 = 1$. 
In both cases we have 
\be
\lambda_{\pm}=\sqrt{\frac{1}{2}\left(-\Lambda\pm\sqrt{\Lambda^2-4\frac{ \tau_{\mathrm{FS}}^2}{L^2}}\right)}\,,
\ee
where
\be
\Lambda=k_{\mathrm{FS}}^2-\frac{1}{L^2}-\tau_{\mathrm{FS}}^2\,.
\ee
Clearly, two space-like separated points on the boundary can be joined via a bulk geodesic isometric to
\be
\gamma^\mu_{\mathrm{Geo}} =
\left(
\begin{array}{c}
0 \\
L\,\tanh(s/L) \\
L\, \sech(s/L)
\end{array} 
\right) \,,
\label{Geo}
\ee
which can be retrieved from either Eq.\,\eqref{Ia} or Eq.\,\eqref{Ic} by setting $\lambda_+=1/L$ . The  curves \eqref{Ia}, \eqref{Ic} and \eqref{Geo} all end on boundary intervals of length $\ell=2L$, every other boundary length can be obtained by a rescaling.

Once the extremal curves have been constructed, we must proceed to evaluate the functional \eqref{torsionaction1}
on these solutions. What makes this functional able to capture the physics of gravitational anomalies is that in contrast to the curve's length, the torsion term is not invariant under boosts \cite{Castro:2014tta}. Observe that the normal vectors in the moving frame \eqref{frame} are defined up to local boosts: the change $n^{A\mu} \to \tilde{n}^{A\mu} = \lambda^A_{\;\;B}[\psi(s)] n^{B\mu}$, with
\be
\lambda^A_{\;\;B}\left[\psi(s)\right]=\left(
\begin{array}{cc}
\cosh\left(\psi(s)\right)&\sinh \left(\psi(s)\right) \\
\sinh\left(\psi(s)\right)&\cosh\left(\psi(s)\right)
\end{array} 
\right)\,,
\ee
leaves the metric $\eta_{AB}$ unvaried.
 Under this transformation, the torsion \eqref{extrinsic curvature} behaves as a gauge connection \cite{Capovilla:1994bs, Carter:2000wv, Fonda:2016ine}
\be\label{gauge}
\tau(s)\rightarrow \tau(s)+\partial_s\psi(s)\,.
\ee
Hence, if the functional \eqref{torsionaction1} is evaluated on an open curve it is not necessarily invariant under gauge transformations, instead it picks up endpoint contributions
\be  
{\cal F}[\gamma] \to {\cal F}[\gamma]+\mathfrak{s}\left[\psi(s_f)-\psi(s_i)\right]\,,
\ee
in a manner analogous to a Chern-Simons action. The crucial consequence of this observation is that to find a unique on-shell value for \eqref{torsionaction1} it is not sufficient to fix boundary conditions for the curve but it is also necessary to impose conditions on the normal vectors. In accordance to \cite{Castro:2014tta} we compel the time-like normal vector to point in a predefined notion of \emph{boundary time} at both endpoints. If instead we were to calculate the on-shell value for a $\kappa-$boosted version of the same curve, while keeping the notion of boundary time fixed, then we would need to adjust the normal frame to satisfy the boundary condition. This adjustment can be implemented by a gauge transformation satisfying $\psi(s)\to-\kappa$ at the curve's endpoints. Therefore, comparing both results we find that under a global Lorentz boost $\Lambda(\kappa)_{\;\;\mu}^\nu$, we have
\be  
{\cal F}\left[\Lambda(\kappa)\gamma\right]-{\cal F}\left[\gamma\right] =-2\,\mathfrak{s}\,\kappa\,,
\ee
which is a manifestation of the quantum violation of boost-invariance: a gravitational anomaly.

We stress that the shape equations \eqref{shape torsion2} are gauge covariant \cite{Fonda:2018eqg}, hence the shape of the Mathisson helix itself is independent of the choice of frame. However, as mentioned before, this is not the case for the on-shell value of the functional \eqref{torsionaction1}. To illustrate this, we consider two different gauge choices: the \textit{Fermi-Walker gauge}, where the torsion is set to zero along the entire curve, and the \textit{Frenet-Serret gauge} where one of the extrinsic curvatures is set to zero identically. In the Fermi-Walker frame, the on-shell value of \eqref{torsionaction1} reads
\be \label{os FW}
{\cal F}_{\mathrm{FW}}[\gamma] =\mathfrak{m}\, \mathrm{L}[\gamma] \,,
\ee
in contrast, for the Frenet-Serret frame we find 
\be \label{os FS} 
{\cal F}_{\mathrm{FS}}[\gamma] =2\mathfrak{m}\, \mathrm{L}[\gamma] \,,
\ee
where we made use of equation \eqref{tau}. Clearly, the Frenet-Serret frame and the Fermi-Walker frame can be related by a gauge transformation. However, this must be a \emph{large} gauge transformation \footnote{See \cite{Banados:2016zim} for an interesting discussion on the subject of gauge transformations and boundary conditions.},
\be
\psi\sim\left( \frac{\mathfrak{m}}{\mathfrak{s}}\right)\,s\,,
\ee
where $s$ is the arc-length parameter. Intuitively, this means that to go from the Frenet-Serret to Fermi-Walker frame we must \emph{unwind} the normal frame an infinite number of times. In both cases, the answer is proportional to the length of the helix $\gamma$, which is given by 
\be\label{helix length}
 \mathrm{L}[\gamma] =\frac{2\,}{\lambda_+}\log\left(\frac{\ell}{\epsilon}\right)\,,
\ee
where $\ell$ is the length of the boundary interval, $\lambda_+$ is a helix parameter and $\epsilon$ is an ultraviolet cutoff.  It is straightforward to check that in in order to obtain the right value for the CFT computation, we must choose the Fermi-Walker frame and $\lambda_+=1/L$ . The latter requirement implies that choosing geodesics as the extremal curve yields the right answer. In view of this fact,  the reader might think that non-geodesic Mathisson's helices play no role in the study of entanglement entropy.
Nonetheless, this would be a premature conclusion as we shall see in the next Section.

\section{Holographic RG flows}

In the previous Section we introduced the shape equations \eqref{shape torsion2} and discussed some of their general properties in AdS$_3$ spacetime. We found that any pair of space-like separated boundary points can be connected using a Mathisson's helix, which could be a geodesic. Now, we consider conformally flat ambient geometries which approach AdS$_3$  asymptotically:
\be\label{interpo}
ds^2=\frac{L_{\mathrm{UV}}^2}{z^2}\left(\eta_{ab}\,dx^adx^b+\frac{dz^2}{f(z)^2}\right)\,,
\ee
with $f(z)\to 1$ as $z\to 0$.
These spacetimes are known to provide a holographic description of the behaviour of renormalization group flows \cite{deBoer:1999tgo}. The infrared ($z\to \infty$) behaviour of these metrics is 
\be
\mathrm{CFT}_{\mathrm{UV}}\to\mathrm{CFT}_{\mathrm{IR}},\;\mathrm{for\;which\;}f(z)\to {L_{\mathrm{UV}}}/{L_{\mathrm{IR}}} \,.
\ee
Spacetimes of this form are studied in detail in \cite{Berg:2001ty}, see also Appendix \ref{app:sam}.
Our task now is to learn how to connect space-like separated boundary points via Mathisson helices.

We view these metrics as solutions to the equations of motion of Topologically Massive Gravity (TMG) \cite{Deser:1982vy} coupled to a scalar field:
\be\label{TMG scalar}
R_{\mu\nu}-\frac{1}{2}g_{\mu\nu} R+\frac{1}{\mu}C_{\mu\nu}=8\partial_\mu\varphi\partial_\nu\varphi-4g_{\mu\nu}\partial^{\lambda}\varphi\partial_\lambda \varphi-\frac{1}{2}g_{\mu\nu}V(\varphi)\,,
\ee
where
\be
C_{\mu\nu}=\epsilon_{\mu}^{\;\lambda\sigma}\nabla_\lambda\left(R_{\sigma\nu}-\frac{1}{4} g_{\sigma\nu}R\right)\,,
\ee
is the Cotton-York tensor. The Cotton-York tensor vanishes for conformally flat geometries, thus, in practice Eqs.\,\eqref{TMG scalar} reduce to Einstein's equations. Nevertheless, we insist on regarding \eqref{interpo} in the context of TMG because the Brown-Henneaux analysis of this theory yields the central charges \cite{Hotta:2008yq}
\be\label{c anom}
c_L=\frac{3L}{2G_3}\left(1-\frac{1}{\mu L}\right)\qquad c_R=\frac{3L}{2G_3}\left(1+\frac{1}{\mu L}\right)\,,
\ee
which signals the presence of the gravitational anomaly. In terms of the TMG couplings, the coefficients of the entangling functional \eqref{torsionaction1} are given by 
\be\label{coupl}
\mathfrak{m}=\frac{1}{4G_3}\qquad \mathfrak{s}=\frac{1}{4G_3\mu}\,,
\ee
 as demonstrated in \cite{Castro:2014tta}. 

For an interpolating geometry of the form \eqref{interpo}, the coupling between the ambient curvature and the moving frame 
reads
\be\label{Curv norm}
R_s^{\;A}=\frac{f'(z)}{zf(z)} n^{A\,z} t^z\,,
\ee
which together with \eqref{change in kFS} implies that $ k^2_{\mathrm{FS}}$ is not necessarily constant along Mathisson helices.  Hence, in contrast to the AdS$_3$ case, it is uncertain whether every geodesic in \eqref{interpo} can be regarded as a Mathisson helix. Indeed, since both extrinsic curvatures $ k^A$ must vanish along geodesics, for them to solve Eq.\,\eqref{shape torsion2} we must have $R_s^{\;A}=0$, which implies that either $t_s^z=0$ or $n^{A\,z}=0$. The first instance corresponds to a curve with a constant $z$ component, that is, a curve lying on a plane parallel to the boundary. In the second case, both normal vectors are orthogonal to the $z$ direction, which implies that the tangent vector itself is orthogonal to the boundary.
Neither of this kind of geodesics can be used to connect space-like separated points in the boundary;
 the former being unable to reach the boundary and the latter touching the boundary only at one point. We conclude that to connect  boundary points we are compelled to use non-geodesic Mathisson helices. Moreover, as these helices approach the boundary they should approximate either \eqref{Ia} or \eqref{Ic}, up to isometries.


\section{Renormalized length-torsion functional}
\label{sec:renorma}

\begin{figure}
  \centering
\includegraphics[scale=.4]{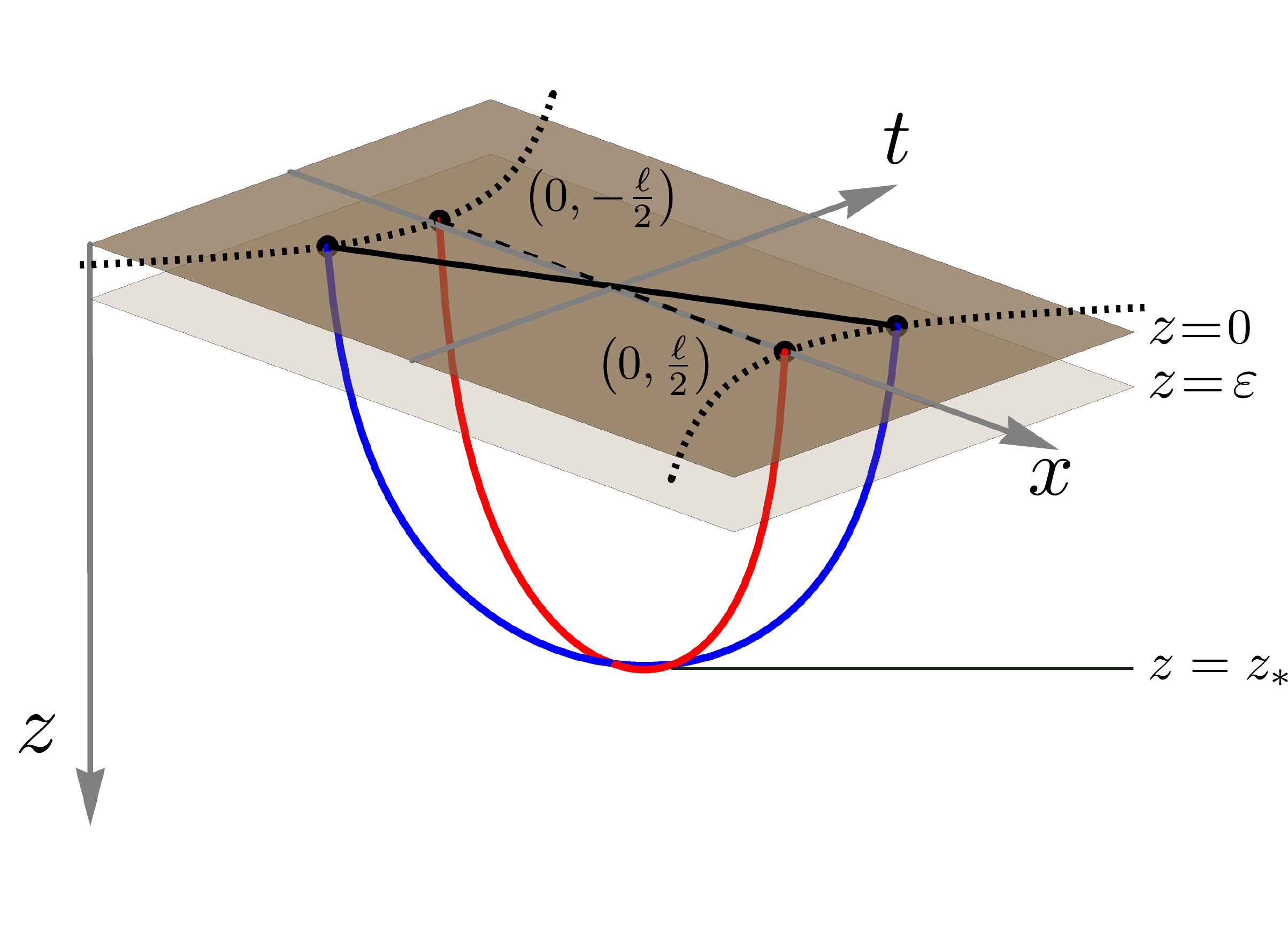}
  \caption{
In the Poincar\'e patch of asymptotically AdS$_3$ spacetimes, we study curves which admit a \textit{tip} $z_*$, i.e. a point where $\dot{z}=0$ and $\ddot{z}<0$. We show two curves with identical $z_*$ but different $\zeta_*$: this corresponds to a rigid boost, hence the endpoints of the red curve lie on the $t=0$ line, while the blue curve endpoints are boosted. In numerical solutions, boundary quantities, such as $\ell$ and the boost $\kappa$, are computed at $z(\pm s_\epsilon)=\epsilon$.
}\label{setup}
\end{figure}

Consider the set of space-like boundary intervals with fixed rapidity $\kappa$ and arbitrary Lorentz-invariant length $\ell$  in the spacetime \eqref{interpo}. Assume that a prescription to construct a unique Mathisson helix $\gamma(\ell)$ connecting the endpoints of each of such intervals has been provided, see Figure \ref{setup}. The question we wish to address is how does the renormalized functional
\be
\widehat{\cal F}\left[\gamma(\ell)\right]=\ell\frac{\partial}{\partial\ell} {\cal F}\left[\gamma(\ell)\right]\,,
\ee
behave as a function of $\ell$. To make progress, it is convenient to express the functional \eqref{torsionaction1} in terms of a Lagrangian density as
\be  \label{integ}
{\cal F}[\gamma(\ell)] =\int_{-s_\epsilon(\ell)}^{s_\epsilon(\ell)}ds\,{\cal L}[\gamma(\ell)]\,,
\ee
with $s_\epsilon(\ell)$ defined by requiring
\be
z\left(\pm s_\epsilon(\ell)\right)=\epsilon\,,
\ee
where $\epsilon$ is an $\ell$-independent ultraviolet cutoff. We denote by $(T_\pm,\,X_\pm)$ the endpoints of the interval, which satisfy
\be\label{EP}
T_+-T_-=\ell\sinh\,\kappa \qquad X_+-X_-=\ell\cosh\,\kappa\,,
\ee
by definition.

The advantage of expressing the functional \eqref{torsionaction1} in the form \eqref{integ} is that the $\ell$-derivative of ${\cal F}[\gamma(\ell)]$ can be separated into two distinct contributions
\be\label{derivative l}
\frac{\partial{\cal F}}{\partial\ell} =\frac{\delta {\cal F}}{\delta \gamma_\mu}\,\frac{\partial \gamma_\mu}{\partial\ell}+\frac{\partial s_\epsilon}{\partial \ell}{\cal L}[\gamma(\ell)]\bigg\rvert_{-s_\epsilon}^{+s_\epsilon}\,.
\ee
The first term follows from the variation of ${\cal F}$ under $\gamma^\mu\to \gamma^\mu+\delta\gamma^\mu$, which can be written as 
\be
\delta {\cal F}[\gamma] =\int_\gamma ds\left[ \,{\cal E}^\mu\delta\gamma_\mu+\partial_s\left({\cal J}^\mu \delta\gamma_\mu\right)\right]\label{variation}\,,
\ee
 the explicit forms of ${\cal E}^\mu$ and ${\cal J}^\mu$ can be found in Eq.\,\eqref{noether 2}, for the present argument only some of their general properties are required. For instance, we will use the fact that $t_\mu{\cal E}^\mu$ vanishes identically and that requiring
 \be
n^A_\mu{\cal E}^\mu=0 
\ee
is equivalent to the shape equations \eqref{shape torsion2}. Hence, since we are considering the variation of the functional about a Mathisson helix, the first term in the right-hand side of \eqref{variation} vanishes. Moreover, given that tangential variations of the functional correspond to reparametrizations of the curve, then
\be\label{Jt}
{\cal J}^\mu t_\mu={\cal L}\,.
\ee
Thus, we can write \eqref{derivative l} as
\be\label{derivative1}
\frac{\partial{\cal F}}{\partial\ell} =\left[{\cal J}_\mu\left(t^\mu\frac{\partial s_\epsilon}{\partial \ell}+\frac{\partial\gamma^\mu}{\partial\ell}\right)\right]_{-s_\epsilon}^{+s_\epsilon}\,.
\ee
Furthermore, the endpoint values of $z$ (which are equal to the cutoff $\epsilon$)  are independent of $\ell$, consequently
\be
\frac{d z}{d\ell}\bigg\rvert_{\pm s_\epsilon}=\left[\dot z\frac{\partial s_\epsilon}{\partial \ell}+\frac{\partial z}{\partial\ell}\right]_{\pm s_\epsilon}=0\,,
\ee
and \eqref{derivative1} becomes
\be\label{derivative2}
\frac{\partial{\cal F}}{\partial\ell} =\left[{\cal J}_\mu\left(\frac{\partial\gamma^\mu}{\partial\ell}-\frac{\dot\gamma^\mu}{\dot z}\frac{\partial z}{\partial \ell}\right)\right]_{-s_\epsilon}^{+s_\epsilon}\,,
\ee
and it is no longer necessary to compute derivatives of $s_\epsilon$.

 Notice that as the helix $\gamma$ reaches towards any of its endpoints, its shape asymptotizes to one of the AdS$_3$ helices described in Sec.\,\ref{sec:AdS hel}. Interestingly, the asymptotic helices approached at each endpoint might be distinct. As shown in Appendix~\ref{asy}, the vector 
\be\label{mira}
\left(\frac{\partial\gamma^\mu}{\partial\ell}-\frac{\dot\gamma^\mu}{\dot z}\frac{\partial z}{\partial \ell}\right)\,,
\ee
evaluated on an AdS$_3$ helix becomes a Killing vector which generates spacetime translations in \eqref{interpo}. Variations taken along any Killing direction $\xi_\mu$ must leave the functional invariant, hence  it follows from Eq.~\eqref{variation} that ${\cal J}^\mu \xi_\mu$  is a conserved quantity. As a matter of fact, in Appendix\,\ref{app:cha} we demonstrate that it matches the spinning-body conserved charge  ${\cal Q}_\xi[\gamma]$ in equation \eqref{conserved}. Bringing these facts together, it follows that the charges are bound to emerge from the contraction inside the bracket in \eqref{derivative2}. Finally, after dealing with a few technical details which can be found in Appendix~\ref{asy}~, we obtain the elegant expression
\be\label{renormal}
\widehat{\cal F}\left[\gamma(\ell)\right]=\ell\left(\,\mathcal{Q}_t[\gamma]\,\sinh\,\kappa +\,\mathcal{Q}_x[\gamma]\,\cosh\,\kappa\right)\,,
\ee
where $\mathcal{Q}_t[\gamma]$ and $\mathcal{Q}_x[\gamma]$ are the Noether charges associated with space and time translations respectively. In particular, setting $\kappa=0$ we have 
\be\label{Myers1}
\widehat{\cal F}\left[\gamma(\ell)\right]=\ell\,\mathcal{Q}_x[\gamma]\,,
\ee
which in the limit $\mathfrak{s}\to 0$ reduces to the entropic c-function constructed in \cite{Myers:2012ed}.
Besides translational symmetries, note that the metric \eqref{interpo} is also endowed with boost invariance. This leads to an additional Noether charge
\be
\mathcal{Q}_b[\gamma] = x \mathcal{Q}_t[\gamma] + t \mathcal{Q}_x[\gamma] + \mathfrak{s} \frac{\dot{z}}{z f(z)} \,.
\ee
Interestingly, for curves that reach the asymptotic AdS boundary, and thus are Mathisson helices of the type discussed in Section \ref{sec:AdS hel}, the last term is always equal to $\mathfrak{s} \lambda_+$. In the following Section we will show that the relevant solutions have $\mathcal{Q}_b[\gamma] =0$ and from the conservation of this quantity it follows that 
\be
\widehat{\cal F}\left[\gamma(\ell)\right]=\mathfrak{s}\lambda_+\left(\frac{\mathcal{Q}_x[\gamma]}{\mathcal{Q}_t[\gamma]}\right)\,,
\label{f hat 2}
\ee
provided $\mathfrak{s} \neq 0$.

\section{Mathisson helices: explicit parametization}

\label{sec:explicit}

In this Section we provide an explicit parametrization to construct the Mathisson helices in an ambient spacetime of the form \eqref{interpo}. In the case of AdS$_3$, the shape equations \eqref{shape torsion2} can be solved following a two-step procedure. First, we solve \eqref{shape torsion2} for the extrinsic quantities $k^A$ and $\tau$ and with these results in hand we then construct the actual curves \cite{Fonda:2018eqg}. Due to the non-trivial coupling of the moving frame with the ambient curvature \eqref{Curv norm}, this procedure cannot be applied in the case of \eqref{interpo}, and we are forced to introduce an explicit parametrization for the curve. As a matter of fact, the best procedure is to introduce a parametrization such that the tangent vector is automatically normalized, so that the resulting curve is parametrized by arc-length. Thus, we introduce functions $\zeta(s)$ and $\delta(s)$ such that the tangent vector reads
\be
\label{tangent}
t^\mu= \frac{z}{L_{\mathrm{UV}}} \left( 
\begin{array}{c}
\sinh \zeta \\
\cos \delta \;\cosh \zeta\\
f(z)\sin \delta \cosh \zeta
\end{array} 
\right)\,.
\ee
Using \eqref{interpo} it is straightforward to verify that $t^\mu t_\mu=1$. Next, we construct the normal frame by inverting Eqs.\,\eqref{frame}. Due to the gauge degeneracy in the normal bundle, we must fix a gauge beforehand in order to find a unique solution. We fix the gauge by demanding that $n^{1\,t}=0$, and obtain
\be
n^{1\mu}=
\frac{z}{L_{\mathrm{UV}}}
\left( 
\begin{array}{c}
0 \\
\sin \delta \\
-f(z) \cos \delta 
\end{array} 
\right)
\qquad
n^{2\mu}=
\frac{z}{L_{\mathrm{UV}}}
\left( 
\begin{array}{c}
 \cosh \zeta  \\
\cos \delta  \sinh \zeta \\
f(z) \sin \delta \sinh \zeta
\end{array} 
\right)\, ,
\ee
for which $\eta_{AB}=\text{diag}(1,-1)$.
In this frame, the curvatures and torsion read
\be
k^1= \frac{f(z)}{L_{\mathrm{UV}}} \cos \delta + \dot{\delta} \cosh \zeta
\qquad
 k^2=\dot{\zeta}-\frac{f(z)}{L_{\mathrm{UV}}} \sin \delta \sinh \zeta
\qquad
\tau=- \dot{\delta}\sinh \zeta \,,
\ee
and the shape equations \eqref{shape torsion2} can be written as
\begin{align}
\label{e1 L}
&
\mathfrak{m}
\left( 
\frac{f(z)}{L_{\mathrm{UV}}} \cos \delta + \dot \delta \cosh \zeta
\right)
+
\mathfrak{s}
\left[
\ddot{\zeta}
+
\cosh\zeta 
\left( 
\dot{\delta}^2\sinh\zeta - \frac{f(z)}{L_{\mathrm{UV}}} \dot{\zeta} \sin \delta
\right)
\right]
=0 \,,
\\
\label{e2 L}&
\mathfrak{m}
\left( 
\frac{f(z)}{L_{\mathrm{UV}}} \sin \delta  \sinh\zeta -\dot \zeta
\right)
+
\mathfrak{s}
\left[
-\cosh \zeta
\ddot{\delta}
+
\dot{\delta}
\left( 
\frac{f(z)}{L_{\mathrm{UV}}} \sin \delta \cosh^2 \zeta - 2 \dot{\zeta} \sinh\zeta
\right)
\right]
=0\,.
\end{align}
These equations, together with the third component of the tangent vector \eqref{tangent} make up a closed system of ordinary differential equations for $\zeta(s)$, $\delta(s)$ and $z(s)$. The $x$ and $t$ coordinates of the curve can be obtained by integrating the respective components of  \eqref{tangent}.

We construct the Mathisson helices by integrating the system of equations presented above using the \emph{shooting method}. We choose a point in the bulk  $\gamma^\mu(0)=(t_*,x_*,z_*)$ where $\dot z(0)=0$, which we call the tip, from which we follow the curve going towards the boundary. Observe that from \eqref{tangent} it follows that $\delta(0)=0$ . Using translational invariance we set $t_*=x_*=0$. Clearly, these conditions are not sufficient to fix a unique solution to the shape equations, additionally we must provide a set of \emph{shooting conditions}:
\be\label{tip cond 1}
\left\{\zeta_* \;, \;\dot\delta_* , \;\dot \zeta* \right\}\,,
\ee
at any depth $z_*$, where the asterisk subscript indicates that the quantity is being evaluated at $s=0$. Alternatively, we can use the values of the extrinsic curvatures 
\be\label{tip cond 2}
\left\{ \zeta_* \;, \;k^1_* , \;k^2_*  \right\}\,,
\ee
as shooting conditions.
The data \eqref{tip cond 1} and \eqref{tip cond 2} can be translated into one another using 
\be\label{shoot2}
k^1_* =\frac{f_*}{L_{\mathrm{UV}}}+\dot\delta_*\cosh\zeta_*\qquad k^2_*=\dot\zeta_* \,.
\ee
In contrast to geodesics where only two parameters, depth and orientation, are needed to fix a unique solution, for Mathisson helices we must provide four. As discussed in Section \,\ref{sec:renorma}, we are interested in helices that connect the endpoints of space-like intervals. It is important to bear in mind that at any given depth $z_*$, only certain choices of shooting conditions will generate a helix that reaches the boundary in this fashion. We say that those conditions belong to the \emph{escape region} of the model. We shall see some examples of these regions in Sec.\ref{sec:numer}. 
 
\begin{figure}
\centering
\includegraphics[width=\linewidth]{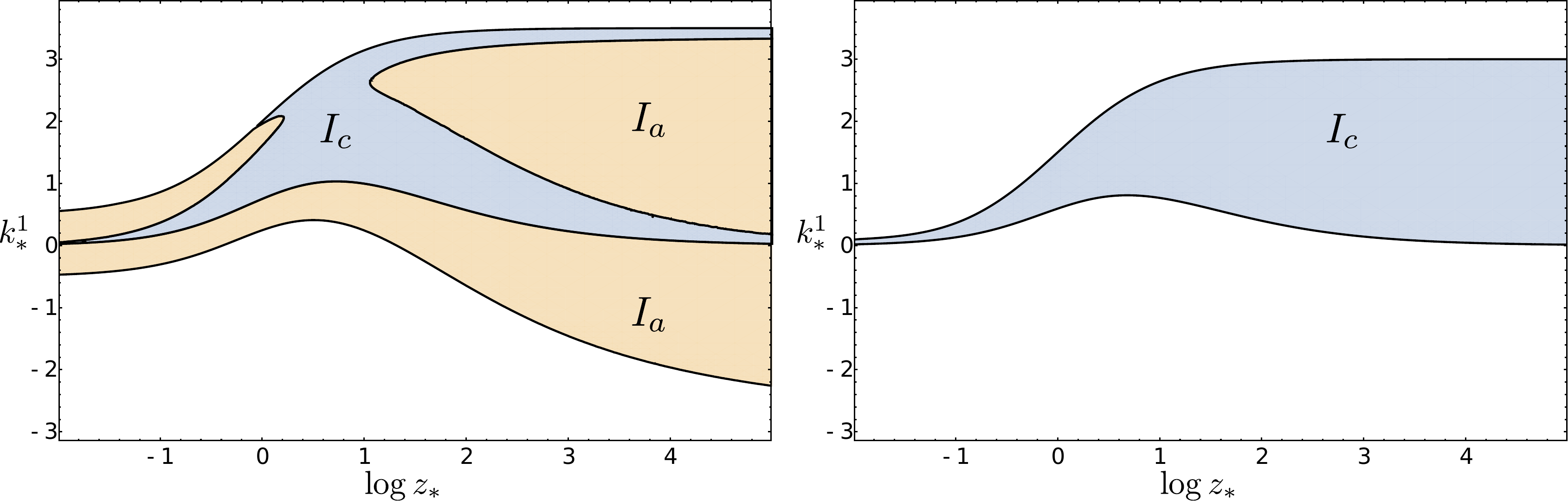}
\caption{Example of escape regions: in the tip parameter space $(z_*, k^1_*)$, we show which values allow for a boundary-reaching solution. The blue/yellow color coding stand for the $I_a$ and $I_c$ helix types, described in \eqref{Ia} and \eqref{Ic}. In both plots we set $L_{UV}/L_{IR}=1/4$, $\mathfrak{m}=1$, $\zeta_*=0$, $\dot{\zeta}_*=0$. We used the interpolating function $f(z)$ in \eqref{f of z}. The left panel has $\mathfrak{s}=2$, while the right one has $\mathfrak{s}=1$. Since for the latter $\mathfrak{m}/\mathfrak{s}=1/L_{UV}$, it represents the chiral point of the theory: notice the absence of type $I_a$ asymptotic behaviour.}\label{escape regions}
\end{figure}

In the parametrization \eqref{tangent}, the spinning-body Noether charges \eqref{conserved} are given by
\bea
{\cal Q}_t &=& 
-\frac{L_{\mathrm{UV}}}{z} \left( \mathfrak{m} \sinh \zeta + \mathfrak{s} \dot \delta \cosh^2 \zeta \right)
\\
{\cal Q}_x &=& 
\frac{L_{\mathrm{UV}}}{z}
\left[ 
\mathfrak{m} 
\cos \delta \cosh \zeta + 
\mathfrak{s}
\left(\dot \delta  \cos \delta \cosh \zeta \sinh \zeta - \dot \zeta  \sin \delta \right)
\right]\,,
\label{QxQt}
\eea
for translational symmetries and
\be
\mathcal{Q}_{b}
=
x \mathcal{Q}_t +
t \mathcal{Q}_x
+\mathfrak{s} \sin \delta \cosh \zeta\,,
\label{Qb}
\ee
for boosts on the $(t,x)$-plane. By evaluating these charges at the tip, we immediately see that $\mathcal{Q}_{b}=0$, while the translational charges can be written as
\be
\label{charges tip}
\left(
\begin{array}{c}
\mathcal{Q}_t \\
\mathcal{Q}_x
\end{array}
\right)
=
\frac{L_{\mathrm{UV}}}{z_*}
\left( 
\begin{array}{cc}
\cosh \zeta_*  & -\sinh \zeta_*\\
-\sinh \zeta_* & \cosh \zeta_*
\end{array}
\right)
\left(
\begin{array}{c}
\displaystyle\mathfrak{s}\left(f_*/L_{\mathrm{UV}}-k^1_*\right)\\ 
\mathfrak{m}
\end{array}
\right)\,.
\ee

\subsection{Helical motions: numerical solutions}
\label{sec:numer}

In this Section we highlight the results from the systematic numerical study of the shape equations performed on backgrounds of the form \eqref{interpo} (and \eqref{samb}). We construct numerical solutions for the system of equations \eqref{tangent}, \eqref{e1 L}, and \eqref{e2 L} parametrized by $\mathfrak{m}$, $\mathfrak{s}$, the tip depth $z_*$, the shooting conditions $\left\{ \zeta_* \;, \;k^1_* , \;k^2_*  \right\}$ and the parameters that determine the warp factor $f(z)$. We produce five interpolating functions:
\begin{align}
\{ t(s),\, x(s),\, z(s),\, \delta(s),\, \zeta(s)\}\,,
\end{align}
and check whether $\gamma$ reaches the asymptotic boundary. Since the spacetime is asymptotically AdS$_3$, the solutions will necessarily approach either a $I_a$ or $I_c$ Mathisson helix, as we showed in Section \ref{sec:AdS hel}. Each solution is unique once we fix initial conditions and parameters. In fact, we can lower the number of degrees of freedom of the solutions by imposing a few extra requirements. The first requirement is that $\gamma$ should approach the \textit{same} type of Mathisson helix on both endpoints, i.e. the solution should have identical values of $\lambda_\pm$ asymptotically. This is equivalent in requiring that the solution should be symmetric under the change of arc-length parameter $s \to -s$, and it is easy to see that this is obtained only if $k^2_*=0$. Some examples of escape regions satisfying this condition are given in Figure \ref{escape regions}. Then, in order for \eqref{renormal} to hold, all curves we consider should also have identical $\kappa$; without loss of generality we can set $\kappa=0$, so both endpoints lie in the $t=0$ slice. Satisfying this condition imposes a non-trivial relationship between the three remaining tip quantities ($\zeta_*$, $z_*$ and $k^1_*$), which we are able to find numerically as a specific value of $\zeta_*$ for fixed $z_*$ and $k^1_*$. Only after this condition is imposed, to each point in the escape region corresponds one and only one Mathisson helix. For every Mathisson helix, we then compute $\ell(z_*,k^1_*)$ and $\widehat{\mathcal{F}}(z_*,k^1_*)$. The final outputs of our algorithm are the values of these quantities within the escape regions; we present some examples in Figures \ref{figure 4} and \ref{figure 5}.

\begin{figure}
\centering
\includegraphics[scale=.3]{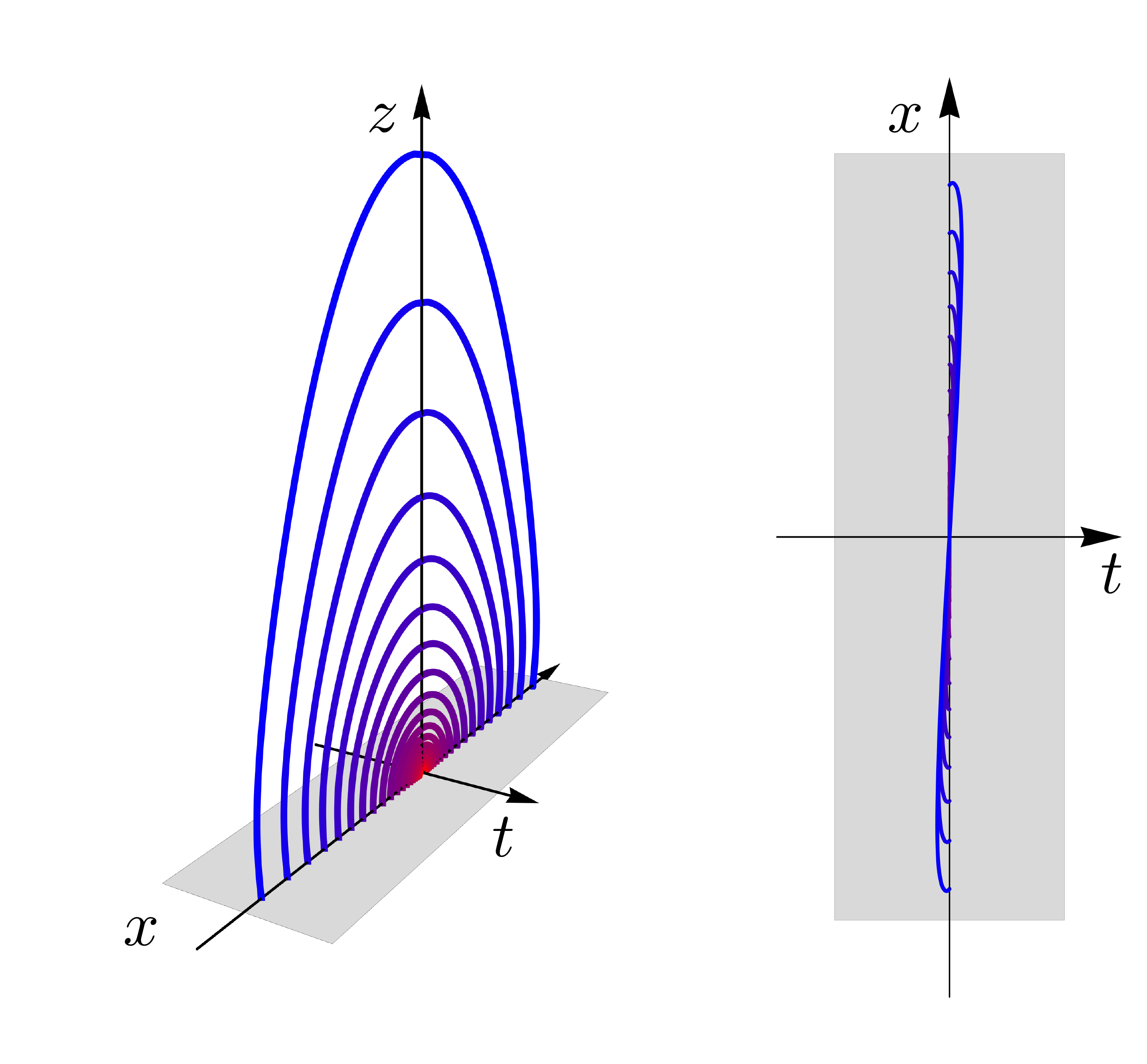}
\caption{Graphical representation of a family of Mathisson helices, solutions of Eqs. \eqref{e1 L} and \eqref{e2 L}. The color of each curve ranges from red to blue as $z_*$ is increased. This family is such that the boundary interval boost, $\kappa$, vanishes (i.e. the endpoints lie on the $t=0$ line) and the asymptotic total curvature $k_{FS}^2$ is zero on both endpoints. These curves are \textit{not} geodesics, and they are non-planar: the vertical view on the right shows this unequivocally. Note that the boundary conditions $\kappa=0$ and $\lambda_+ =1$ can be reached only with a specific choice of tip values $\zeta_*$ and $k_*^1$. This choice changes non-trivially for different $z_*$.}\label{curves}
\end{figure}

\begin{figure}[t]
\centering
\includegraphics[width=\linewidth]{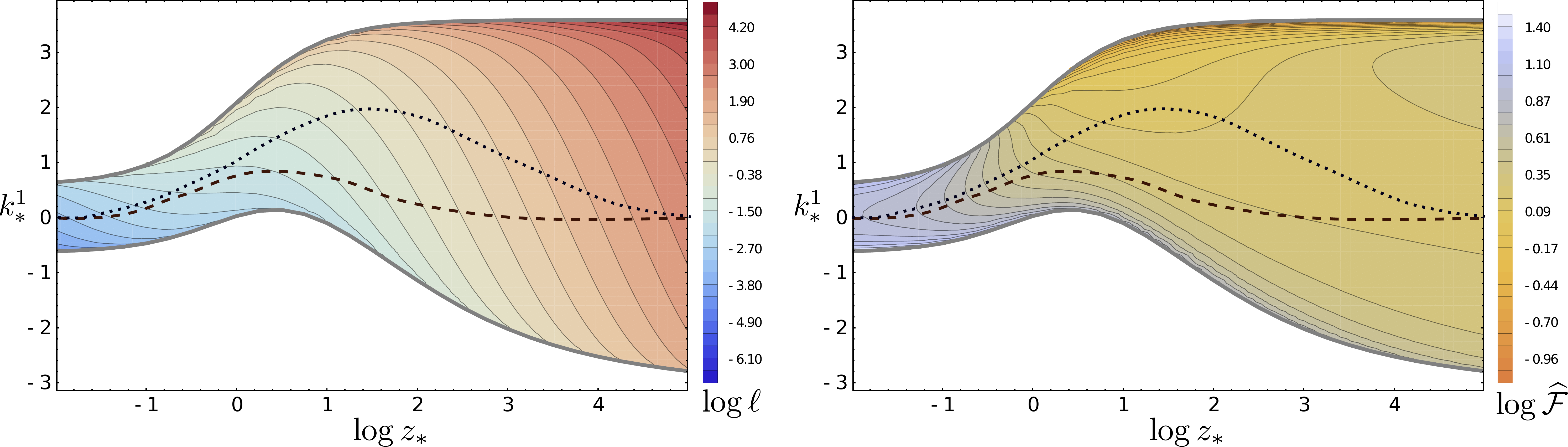}
\caption{We take the escape region in the left panel of Figure \ref{escape regions}, and plot logarithmic contour lines $\ell(z_*,k^1_*)$ (on the left) and $\widehat{\mathcal{F}}(z_*,k^1_*)$  (on the right). For each point within the region, we carefully tuned $\zeta_*$ so that each Mathisson helix ends at the boundary with $\kappa=0$, as in Figure \ref{curves}. Within the escape region we also show two black lines, as exemplification of possible choices for curve prescriptions. The dashed (lower) black line correspond to a prescription which, as we change $\ell$ is also monotonic in $\widehat{\mathcal{F}}$. On the other hand, the dotted black line is an unsuitable choice because for increasing $\ell$, the value of $\widehat{\mathcal{F}}$ decreases (this happens where the line crosses the same contour twice).}\label{figure 4}
\end{figure}

\begin{figure}[t]
\centering
\includegraphics[width=\linewidth]{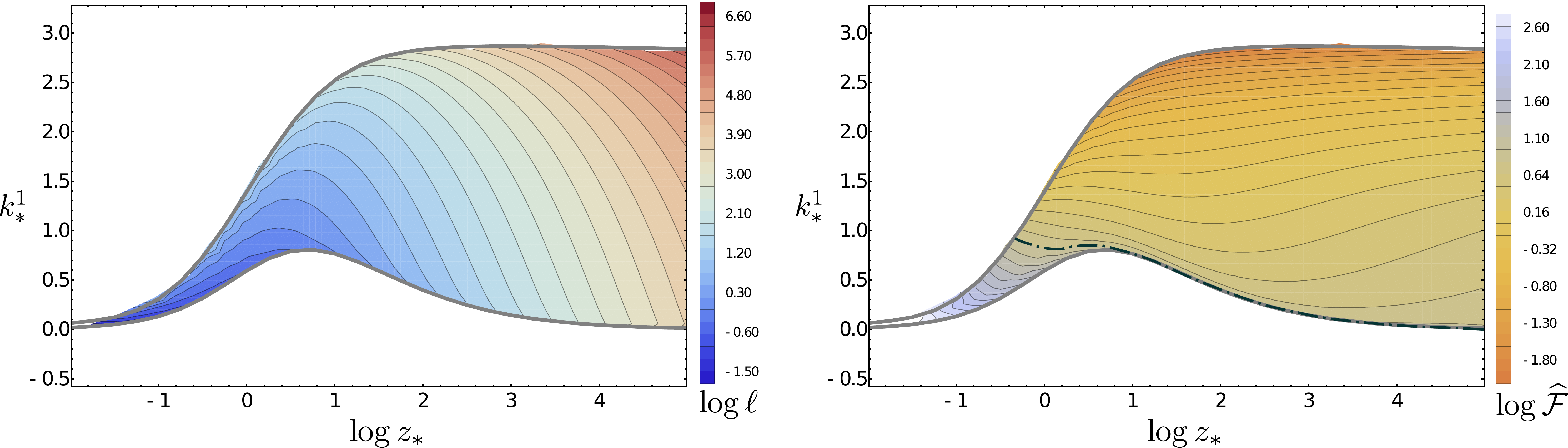}
\caption{
The escape region at the chiral point, taken from the right panel of Figure \ref{escape regions}, with logarithmic contour lines of $\ell(z_*,k^1_*)$ (on the left) and $\widehat{\mathcal{F}}(z_*,k^1_*)$  (on the right). The vertical axis has been rescaled w.r.t. Figure \ref{escape regions}. The dot-dashed black line in the right panel shows the critical contour discussed in the main text. The value of $\widehat{\mathcal{F}}$ along this line is precisely the infrared CFT value of the central charge: therefore, monotonicity requires that any asymptotic prescription must lie below this limiting contour.}\label{figure 5}
\end{figure}

Based on these findings we would like to construct \textit{entropic} $c-$functions out of Mathisson's helices. Attaining this involves selecting a suitable family of Mathisson helices; in practice this correspond to delineating a properly chosen trajectory within the $(z_*,k^1_*)$ plane. First of all, we require that this trajectory relates monotonically the depth $z_*$ with the boundary width $\ell$, which yields a family of curves as the one depicted in Fig.~\ref{curves}. Furthermore, we demand this trajectory to reproduce the expected CFT values 
\be
c_{\mathrm{UV}}=6\,\mathfrak{m}\,L_{\mathrm{UV}}\qquad \text{and}\qquad c_{\mathrm{IR}}=6\,\mathfrak{m}\,L_{\mathrm{IR}}\,
\ee
as $z_*\to 0$ and $z_*\to \infty$, respectively, and to interpolate between these monotonically. This means that we must select a trajectory with $k^1_*\to 0$ in the UV as well as at the IR. This trajectory must cross each contour in both plots in Fig.~\ref{figure 4} once and only once. In general there are infinitely many ways of attaining this. However, at the chiral point, where $\mathfrak{m}/\mathfrak{s}=1/L_{UV}$, the contours behave in a different manner (see Fig.~\ref{figure 5}): they either intersect the lower boundary of the escape region or continue freely towards the IR, the critical value between these two cases precisely being the contour corresponding the the expected CFT value in the IR. Clearly, monotonicity requires that our trajectory remains below that contour. From our numerical results, we notice that as we increase the ratio $   L_{\mathrm{UV}}/L_{\mathrm{IR}}$, the critical contour comes closer to the lower boundary of the escape region. Therefore, if we want to have a general prescription we must make our trajectory match the lower boundary of the escape region; this corresponds to a family of asymptotically geodesic Mathisson helices. While this choice is clearly a necessary condition for monotonicity it does not guarantee it. However, we
also considered other geometries in Appendix \ref{app:sam}, and obtained qualitatively similar results (compare Figure \ref{figure 5} with Figure \ref{figure 6} in the Appendix). Thus, we propose an entropic $c$-function
\be\label{helical c}
\frac{c_{\mathrm{Hel}}(\ell)}{3} =\frac{\mathcal{Q}_x}{\mathcal{Q}_t}  \,,
\ee
at the chiral point based on the renormalized EE functional \eqref{torsionaction1} and which can be computed entirely in terms of spinning body conserved charges. Even though the curves used to compute \eqref{helical c} are asymptotically geodesic, it is important to point out that this quantity is different from the non-anomalous case ($\mathfrak{s}=0$). In fact, using \eqref{Myers1} and the charges \eqref{charges tip}, we can write
\be
\widehat{\cal F} =L_{\mathrm{UV}}\left(\frac{\ell}{z_*}\right)\left[\mathfrak{s}\sinh(\zeta_*)\left(k^1_*-f_*/L_{\mathrm{UV}}\right)+\mathfrak{m}\cosh(\zeta_*)\right]\,.
\ee
In the non-anomalous case, extremal curves are geodesics with $\zeta_*=0$ hence
\be\label{Myers2}
\widehat{\cal F} =\frac{c_{\mathrm{Geo}}(\ell)}{3}=\mathfrak{m}L_{\mathrm{UV}}\left(\frac{\ell}{z_*}\right)\,,
\ee
which is precisely the result found in \cite{Myers:2012ed}.

\section{Discussion and outlook}

In this work we have explored the physical properties of the length-torsion functional \eqref{torsionaction1}. As demonstrated in \cite{Castro:2014tta}, this functional computes holographically the entanglement entropy for 1+1 dimensional theories with chiral anomalies. Moreover, in a previous paper \cite{Fonda:2018eqg} we have shown that the extremal curves of this functional correspond to Mathisson's helical motions for the centers of mass of spinning bodies. Here, we have brought together these two points of view and constructed an entropic $c$-function $c_{\mathrm{Hel}}(\ell)$ which can be written in terms of Noether charges along Mathisson's helices. While for generic values of the anomaly
there is some ambiguity in the definition of $c_{\mathrm{Hel}}(\ell)$, we argue that at the chiral point \eqref{chira} this ambiguity  is absent: we find the succinct expression \eqref{helical c}, which must be evaluated on asymptotically geodesic Mathisson helices. While we have gathered extensive numerical evidence in support of the monotonicity of this function, we leave the derivation of a formal proof of this fact for future work. 

We wish to point out that the steps leading to the expressions for the renormalized functional, Eqs.\eqref{renormal} and \eqref{f hat 2}, relied only on the assumption of having an AdS$_3$ asymptopia. Hence, these expressions can be used without any modification for any IR behaviour one might wish to study, such as warped AdS$_3$, Janus or gapped geometries. Indeed, for the case of gapped geometries we will detail in a forthcoming work how Mathisson helices prevent the formation of mass gaps which are precluded by anomaly matching, see \cite{Belin:2015jpa} for a general argument. Another intriguing direction is to revisit entanglement entropy for theories with different symmetry algebras and whose gravitational duals are naturally understood in the realm of TMG such as Galilean CFTs, see for instance \cite{Bagchi:2014iea} and \cite{Hosseini:2015uba}. On a more ludic note, when exploring the space of Mathisson helices in domain walls, we noticed that beyond the escape regions of Figure \ref{escape regions}, there are also extra, small regions in the parameter space which correspond to a curious kind of solutions: curves in these regions escape towards the boundary only after travelling deeper into the bulk and gathering enough momentum from the spin-curvature interaction. The tip $z_*$, for these curves, is only a local maximum of $z(s)$. It would be amusing to explore what these solutions might be able to teach us. Finally, we wish to understand the results discussed in this work from a field theoretic standpoint.  First, it would be desirable to obtain a CFT picture of Mathisson's helices and their charges. Based on this, we ought to be able to translate $c_{\mathrm{Hel}}(\ell)$ into the language of the dual theory. Presumably, this would allow us to make contact with the very interesting literature concerning EE in CFTs with chiral anomalies, \cite{Azeyanagi:2015gqa,Nishioka:2015uka,Iqbal:2015vka,Hughes:2015ora}.

\section*{Acknowledgements}

PF is supported by the Netherlands Organisation for Scientific Research (NWO/OCW).
The work of AVO is supported by the NCN grant 2012/06/A/ST2/00396. AVO wishes to thank the Yukawa Institute for Theoretical Physics for hospitality during the development of this work.
Also,  AVO  is  grateful  to  Abdus Salam International Center for Theoretical Physics where he carried out some of the final stages of this project. It is a pleasure to acknowledge Michael Abbott, Pawe\l{} Caputa, Alejandra Castro, Filipe Costa, Mario Flory, Vishnu Jejjala and Jos\'e Natario for enlightening conversations and correspondence. We thank Alejandra Castro, Mario Flory, Vishnu Jejjala and Hesam Soltanpanahi for helpful comments on earlier versions of this work. Especially, we wish to express our gratitude to Alejandra Castro for suggesting us to look into these questions. 

\appendix

\section{Helical Noether charges}
\label{app:noether}

In this Appendix we show how the quantity \eqref{conserved} can be derived from the action \eqref{torsionaction1} by means of Noether's theorem. 

Since $\{t^\mu,n^{1\mu},n^{2\mu}\}$ form an oriented basis frame in the neighbourhood of the curve, we can decompose any (Killing) vector field $\xi^\mu$ defined on $T M$ in its components. Explicitly we have
\be
\xi^\mu = \xi_t t^\mu + \xi_A n^{A\mu} \,,
\label{Killing proj}
\ee
where $\xi_t$ and $\xi_A$ are respectively the tangential and normal components of the Killing vector field. 

Note that the expression written in \eqref{conserved} is gauge invariant, i.e. does not depend on any particular choice of the curve's normal frame. For this reason, it is convenient to (temporarily) make \eqref{torsionaction1} also gauge invariant by adding a \textit{compensator}:
\be  
\tilde{\mathcal{F}}[\gamma] \equiv  \int_\Sigma ds \left(\mathfrak{m} + \mathfrak{s} \left( \tau - \dot{\psi}\right) \right) \,,
\label{compensator}
\ee
where $\psi(s)$ is the hyperbolic angle of $n^{1\mu}$ with any arbitrarily fixed normal frame. For example, we can choose $\psi$ to be the angle with the Fermi-Walker gauge choice, so that \textit{if and only if} we compute geometrical quantities in this frame, we can set $\psi =0$. Under a local frame rotation the compensator term changes in exactly the opposite way as $\tau$, rendering \eqref{compensator} effectively gauge invariant.
The compensator arises also in the evaluation of some connection forms, namely 
\be 
\frac{1}{2} \epsilon_{AB} n^{A\mu} n^{B\nu} \nabla_\mu t_{\nu} = \tau - \dot{\psi} \,,
\ee
being the l.h.s of this expression a well-behaved scalar. 

We can now substitute \eqref{Killing proj} into \eqref{conserved}, finding
\bea
\mathcal{Q}_\xi[\gamma]
&=&
\xi_t 
\left(\mathfrak{m} + \mathfrak{s} \left( \tau - \dot{\psi}\right) \right)
-
\mathfrak{s} 
\epsilon_{AB}
\left( 
\xi^A \tr K^B
+
\frac{1}{2} n^{A\mu} \nabla_\mu \xi^B 
+
\frac{1}{2} \xi_C \Theta^{ACB}
\right) \,.
\label{noether 1}
\eea
On the other hand, by taking an on-shell Lie derivative of \eqref{compensator} along the vector field \eqref{Killing proj}, we get
\be
\mathcal{L}_\xi \tilde{\mathcal{F}}[\gamma] 
= 
\left[ 
\xi_t 
\left(\mathfrak{m} + \mathfrak{s} \left( \tau - \dot{\psi}\right) \right)
-
\mathfrak{s} 
\epsilon_{AB}
\left( 
\xi^A \tr K^B 
-
\frac{1}{2} \xi_C \Theta^{CBA}
\right)
\right]_{s=\pm \infty}\,.
\label{noether 2}
\ee
To obtain the above result we used the fact that the bulk of the variation is zero because of the shape equations, and used the fact that $\mathcal{L}_\xi \psi=0$ at the boundary. Because of Noether's theorem, if $\xi^\mu$ is a Killing vector field, being \eqref{compensator} a geometrical invariant action (i.e. it does not depend on coordinate choices), then the term inside the brackets of \eqref{noether 2} and expression \eqref{noether 1} should match.

In order to prove the equivalence of the two expressions, we need to use the fact that Killing vectors preserve orthonormal frames. In particular, the normal vectors can be Lie-transported along any spacetime Killing directions
\be
\mathcal{L}_\xi n^{B\mu} = \xi^\nu \nabla_\nu n^{B\mu} - n^{B\nu} \nabla_\nu \xi^\mu = 0 \,. 
\ee
By contracting the above expression with $\epsilon_{AB} n^{A}_\nu$ we get the relation
\be
\epsilon_{AB} \left( \xi_C \Theta^{CAB} - n^{A\mu}\nabla_\mu \xi^B - \xi_C \Theta^{ACB} \right) = 0 \,,
\ee
which, upon using $\xi_C \Theta^{C[AB]}=0$, proves the equivalence between \eqref{noether 1} and 
\eqref{noether 2}.

\section{Spinning-body charges in AdS$_3$}
\label{app:cha}

In the following, we compute the conserved charges of helices isometric to \eqref{Ia} or \eqref{Ic} in AdS$_3$ (here, the AdS radius is equal to $L$). Specifically, we are interested in helices of the form
\be
\label{apphelx}
\gamma^\mu(s) = \frac{r}{2 L}M^\mu_{\;\;\nu} x^\nu(s) + \Delta^\mu,
\ee
where $x^\nu$ is either $\gamma^\mu_{I_a}$ (see \eqref{Ia}) or $\gamma^\mu_{I_c}$ (see \eqref{Ic}). The coefficient $r/2 L_{UV}$ sets the boundary interval length to $\ell = r$, while $M^\mu_\nu$ and $\Delta^\mu$ are respectively a rigid boost and a translation
\be
M^\mu_{\;\;\nu} = \left(
\begin{array}{ccc}
 \text{cosh} \rap  & -\text{sinh} \rap &0  \\
 -\text{sinh} \rap  & \text{cosh} \rap &0  \\
 0&0&1\\
\end{array}
\right)\,, \qquad \Delta^\mu= \left(\begin{array}{c} \Delta^t\\ \Delta^x\\ 0\end{array}\right).
\ee
A simple approach to compute the Noether charges onto \eqref{apphelx} is to regard the ambient space as a hypersurface embedded in a four-dimensional flat space described as the zero set of $y^\alpha y_\alpha + L^2$ in $\mathbb{R}^4$, endowed with the metric diag$(-,-,+,+)$. The map we use for the embedding is
\be
(t,x,z) = \frac{L}{y^1+y^4}\left(y^2,y^3,L\right)
\ee 
with inverse mapping 
\be
\label{impmap}
(y^1,y^2,y^3,y^4) = \frac{1}{2z}\left(L^2-t^2+x^2+z^2, \; 2Lt, \; 2Lx,\; L^2+t^2-x^2-z^2\right).
\ee
In these coordinates, the Killing vectors in $\mathbb{R}^{2,2}$ associated to translations and boost are respectively
\bea
\xi_t^\mu &=& \frac{1}{L}\left(-y^2,y^1+y^4,0,y^2\right)
\,,\\
\xi_x^\mu &=& \frac{1}{L}\left(y^3,0,y^1+y^4,-y^3\right)
\,,\\
\xi_b^\mu &=& \frac{1}{L}\left(0,y^3,y^2,0\right).
\eea
To compute ${\cal Q}_t$ and ${\cal Q}_x$ we use the FS frame to first calculate the momentum $p^\mu$ and spin $S^{\mu\nu}$ of the curve. This is a rather complicated task using the the Poincar\'e coordinates for AdS$_3$, the main difficulty arising from the fact that ${\cal D}_s$, defined in \eqref{extrinsic curvature}, is not a ordinary derivative. To circumvent this complication, we compute these quantities in the four-dimensional spacetime. Namely, in $\mathbb{R}^{2,2}$ the FS frame equations are linear due to the fact that the directional derivative is simply by 
\be
{\cal D}_s V^\alpha = \partial_s V^\alpha + \frac{1}{L_{\mathrm{UV}}^2}(y_\beta \partial_s V^\beta)y^\alpha \,.
\ee
This linear realization of the FS equations allowed us to construct all types of hyperbolic helices in \cite{Fonda:2016ine}.

To further simplify the computations, we can momentarily take $\Delta^\mu$ to be zero, since ${\cal Q}_t$ and ${\cal Q}_x$ cannot depend on the absolute position in the $(t,x)$ plane of the curve.  We find
\be
\label{apQxQt}
\left(
\begin{array}{c}
{\cal Q}_t \\
{\cal Q}_x
\end{array}
\right) 
= 
\frac{2 L}{r}
\left(
\begin{array}{cc} 
\cosh\rap & \sinh\rap \\
\sinh\rap & \cosh\rap
\end{array}
\right)
\left( 
\begin{array}{c} 
\mathfrak{s}\lambda_+ \\
\mathfrak{s}\lambda_- 
\end{array}
\right)\,.
\ee
To compute ${\cal Q}_b$ we make use of the fact that
\be
\frac{L\;t^z}{z f(z)} = \sin \delta \cosh\zeta \,,
\ee
from which we find, for arbitrary translation parameters $\Delta^\mu$, that
\be
{\cal Q}_b = {\cal Q}_x\Delta^t + {\cal Q}_t \Delta^x \,.
\ee
Since $Q_b=0$, this equation implies the relation
\be
{\cal Q}_t \Delta^x = - {\cal Q}_x \Delta^t \,.
\label{QDelta}
\ee

\section{Explicit Holographic RG flow geometries}
\label{app:sam}

\begin{figure}[th!]
\centering
\includegraphics[width=.95\linewidth]{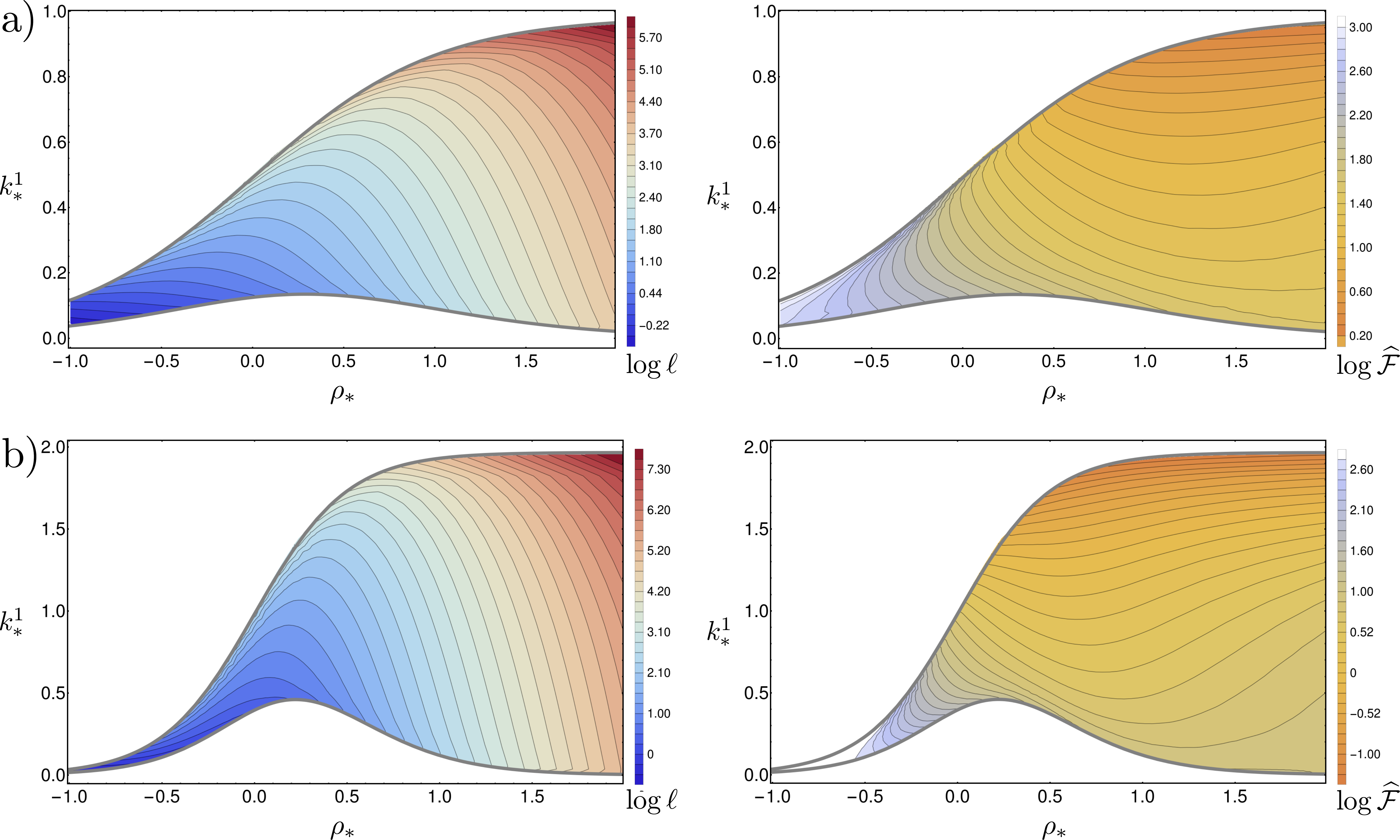}
\includegraphics[width=.95\linewidth]{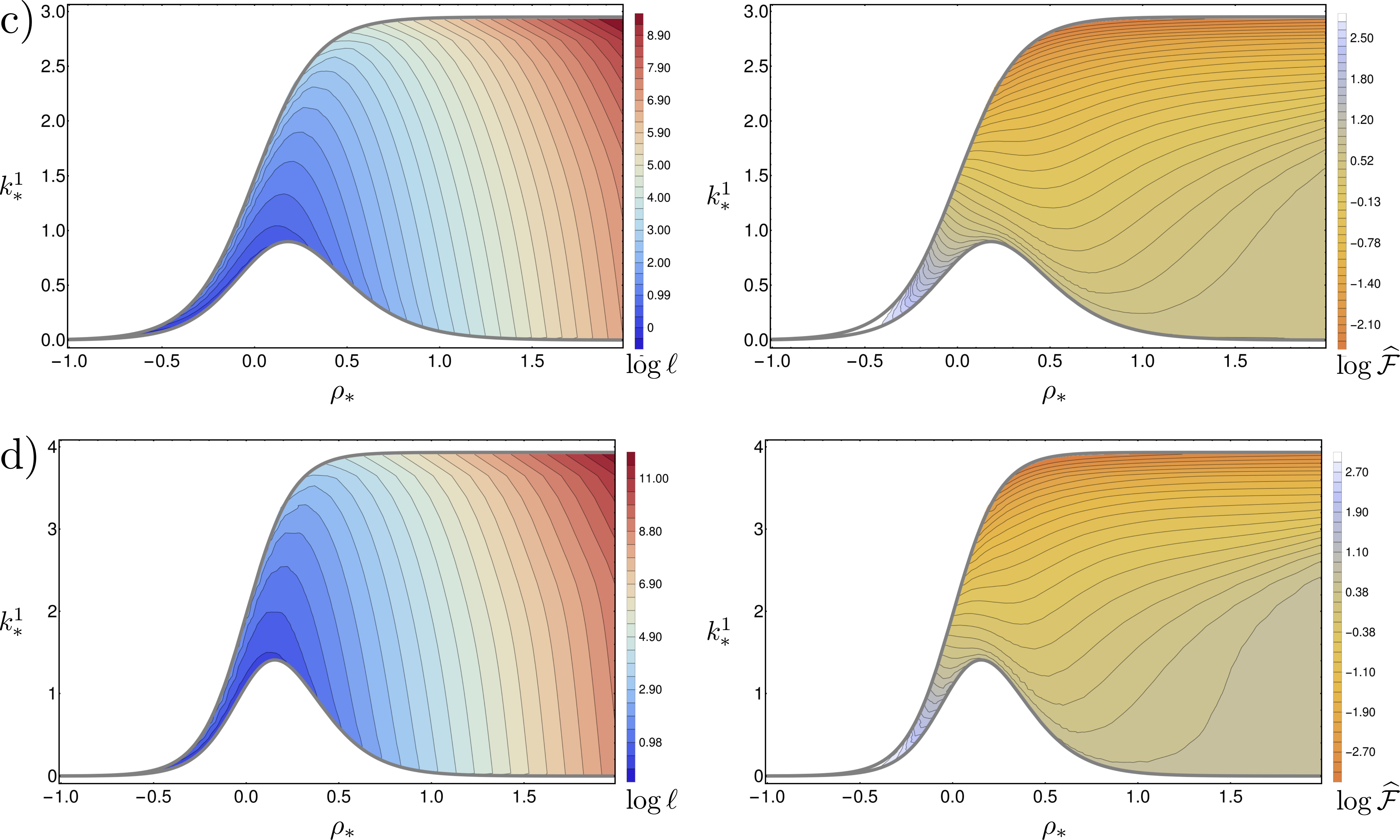}
\caption{As in Figure \ref{figure 6}, we show the contour plots of $\ell$ and $\widehat{\mathcal{F}}$ within the escape regions at the chiral point, now for the domain wall metric \ref{samb}. From panels a) to d) we increase $L_{UV}/L_{IR}$ from $2$ to $5$. In all cases there is a critical contour: the only commonly valid prescription for the c-function is to choose asymptotically geodesic helices.
}\label{figure 6}
\end{figure}

In the main text we make use of RG flow geometries of the form \eqref{interpo}, where we did not need to specify the functional for of the interpolating function $f(z)$. However, for all numerical purposes it is necessary to choose a specific function. In all solutions involved in Figures \ref{curves}, \ref{figure 4} and \ref{figure 5} we made the explicit choice
\be
f(z) 
= \frac{1+\frac{L_{\mathrm{UV}}}{L_{\mathrm{IR}}}z^2}{1+z^2} \,,
\label{f of z}
\ee
which has the key property of interpolating between $1$ for $z\to 0$ and $\frac{L_{\mathrm{UV}}}{L_{\mathrm{IR}}}$ for $z \to +\infty$. Although this $f(z)$ is monotonic, the function \eqref{f of z}  is only a \textit{phenomenological} choice, since it does not descend from any \textit{top-down} model.

For this reason, and also to test our analytical results with other geometries, we repeat the numerical analysis of Section \ref{sec:numer} using a second type of metrics: the analytic domain wall solution of three-dimensional $SO(4)\times SO(4)$ gauged supergravity constructed in \cite{Berg:2001ty}. In these solutions, the three-dimensional metrics are all of the form
\be\label{samb}
ds^2 = e^{-2A(\rho)}\eta_{ab} dx^a dx^b+ d\rho^2\,,
\ee
with
\be
A(\rho) = \frac{\alpha}{2 \beta}\rho-\frac{1}{2}\log\left[\textnormal{sech}\left(\frac{\rho}{\beta}\right)\right] \,,
\ee
where
\be
\quad \alpha = \frac{L_{\mathrm{IR}}+L_{\mathrm{UV}}}{L_{\mathrm{UV}}-L_{\mathrm{IR}}},\quad \beta = \frac{L_{\mathrm{IR}} L_{\mathrm{UV}}}{L_{\mathrm{UV}}-L_{\mathrm{IR}}}\,.
\ee
Since this geometry is given in a different coordinate patch than Eq.~\eqref{interpo}, for the reader's convenience we reproduce below some of the basic formulas of Section \ref{sec:explicit}. Some numerical results are shown in Figure \ref{figure 6}.

We use the following arc-length parametrization of the tangent vector
\be
t^\mu = 
\begin{pmatrix}
e^{A(\rho)} \sinh\zeta \\
e^{A(\rho)}\cos\delta\cosh\zeta \\
\sin\delta\cosh\zeta
\end{pmatrix} 
\,,
\ee
with orthogonal frame
\be
n^{1\mu} = 
\begin{pmatrix}
0 \\
e^{A(\rho)}\sin\delta \\
-\cos\delta
\end{pmatrix}
\,, \qquad
n^{2\mu}
= 
\begin{pmatrix}
e^{A(\rho)}\cosh\zeta \\
e^{A(\rho)}\cos\delta\sinh\zeta \\
\sin\delta\sinh\zeta
\end{pmatrix}
\,.
\ee
The associated curvatures and torsion are
\be
k^1 = A'(\rho)\cos\delta +\dot \delta \cosh\zeta\,, \qquad 
k^2 = \dot\zeta -A'(\rho)\sin\delta \sinh\zeta\,, \qquad 
\tau = -\dot \delta \sinh\zeta \,.
\ee
The shape equations are given by
\bea
\mathfrak{m}
\left( A'(\rho) \cos\delta + \dot \delta \cosh\zeta\right) 
+ 
\mathfrak{s}
\left[\ddot \zeta + \cosh\zeta \left(\dot\delta^2\sinh\zeta-A'(\rho)\dot \zeta \sin\delta\right)\right]
 = 0 \,, \\
\mathfrak{m}
\left( A'(\rho)\sin\delta\sinh\zeta - \dot \zeta\right)
+
\mathfrak{s}
\left[ - \cosh\zeta \ddot \delta +\dot \delta \right(A'(\rho)\sin\delta\cosh^2\zeta-2\dot \zeta\sinh\zeta\left)\right]
= 0 \,.
\eea
Not surprisingly, all of the above expressions can be obtained from their analogues of Section \ref{sec:explicit} by replacing $f(z)/L_{\mathrm{UV}} \rightarrow A'(\rho)$. By imposing the tip-defining condition $\dot\rho(0)=0$, with fixed values for $\rho_*, \zeta_*$, $k^1_*$ and $k^2_*$, we get 
\be
\delta_* = 0
\,,\qquad
\dot\delta_* = (k^1_*-A'(\rho_*))\,\sech \zeta_* 
\,,\qquad 
\dot\zeta_* = k^2_* \,.
\ee
The Noether charges are given by
\bea
{\cal Q}_t &=& 
-e^{-A(\rho)}\left(\mathfrak{m}\sinh\zeta + \mathfrak{s}\dot\delta\cosh^2\zeta\right) \,,\\
{\cal Q}_x &=& 
e^{-A(\rho)}\left[\mathfrak{m}\cos\delta\cosh\zeta+\mathfrak{s}\left(\dot\delta\cos\delta\cosh\zeta\sinh\zeta-\dot\zeta\sin\delta\right)\right] \,,\\
{\cal Q}_b &=& x {\cal Q}_t + t{\cal Q}_x +  \sigma \cosh\zeta \sin\delta \,.
\eea
Again, the tip condition imposes ${\cal Q}_b=0$. The two remaining charges are instead
Evaluating the charges at the tip, they can be written as:
\be
\begin{pmatrix}
{\cal Q}_t \\ 
{\cal Q}_x
\end{pmatrix} 
= 
e^{-A_*} 
\begin{pmatrix}
\cosh\zeta_* & -\sinh\zeta_* \\
-\sinh\zeta_* & \cosh\zeta_*
\end{pmatrix}
\begin{pmatrix}
\mathfrak{s}(A'_*-k^1_*) \\
\mathfrak{m}
\end{pmatrix} \,.
\ee

\section{Matching charges}
\label{asy}

In this Section we connect the Mathisson helices behaviour at the boundary with their tip values, eventually with the objective of proving equation \eqref{renormal}. To this purpose, we exploit the fact that solutions to the shape equations \eqref{shape torsion2} are asymptotically AdS$_3$ helices of the form \eqref{apphelx}. Since each helix might have a different asymptotic behaviour on either of its boundary endpoints, we will supply the notation of $(r, \rap, \Delta^t, \Delta^x)$ of Appendix \ref{app:cha} with a $(\pm)$ superscript to indicate whether they refer to the $s\to \pm \infty$ limits. The conserved charges ${\cal Q}_t$, ${\cal Q}_x$ can be expressed in two different ways, using tip (see \eqref{QxQt}) and boundary (see \eqref{apQxQt}) quantities. By comparing these two expressions, we can solvefor the dilatations $r^{(\pm)}$ and boost $\rap^{(\pm)}$ boundary parameters, finding
\bea
\label{rapp1}
r^{(\pm)} &=& 
2 \mathfrak{s} z_* \sqrt{\frac{\left(\lambda_+^{(\pm)}\right)^2-\left(\lambda_-^{(\pm)}\right)^2}{\mathfrak{s}^2 \dot \delta_*^2 \cosh(\zeta_*)^2-\mathfrak{m}^2}} \,,\\
\label{rapapp1}
\rap^{(\pm)} &=& 
-\zeta_*
+
\frac{1}{2}
\log
\left[
\frac
{(\mathfrak{s}\dot \delta_* \cosh\zeta_*-\mathfrak{m})\left(\lambda_+^{(\pm)}-\lambda_-^{(\pm)}\right)}
{(\mathfrak{s}\dot \delta_* \cosh\zeta_* + \mathfrak{m})\left(\lambda_+^{(\pm)}+\lambda_-^{(\pm)}\right)}
\right] \,.
\eea
Since ${\cal Q}_b=0$ we have that \eqref{QDelta} holds on both endpoints, i.e.
\be
\label{appAeqQr}
{\cal Q}_x \Delta^{t(\pm)} + {\cal Q}_t \Delta^{x(\pm)} =0 \,.
\ee

The significance of these relations relies in the fact that we can use them to find the length $\ell$ and the rapidity $\kappa$ of the interval bounded by endpoints $(T_\pm,X_\pm)$. For helices isometric to $\gamma^\mu_{I_a}$ or $\gamma_{I_c}^\mu$, we find that
\bea
\label{relapp1}
T_{\pm} &=& \Delta^{t(\pm)} \mp r^{(\pm)}\sinh(\rap^{(\pm)}) \,,\\
\label{relapp2}
X_{\pm} &=& \Delta^{x(\pm)} \pm r^{(\pm)}\cosh(\rap^{(\pm)}) \,.
\eea
Using these relations and equation \eqref{EP} we get
\bea
\label{appAeq1}
\Delta^{t(+)}- \Delta^{t(-)}
&=& 
\ell \sinh\kappa + \frac{1}{2}\left(r^{(+)} \sinh\rap^{(+)} + r^{(-)}\sinh\rap^{(-)}\right) \,,
\\
\label{appAeq2} 
\Delta^{x(+)}- \Delta^{x(-)}
&=& 
\ell \cosh\kappa - \frac{1}{2}\left(r^{(+)} \cosh\rap^{(+)} + r^{(-)}\cosh\rap^{(-)}\right) \,.
\eea
By substituting the relations \eqref{appAeq1} and \eqref{appAeq2} into \eqref{mira} we find that
\be
\left[ \frac{\partial \gamma^\mu}{\partial \ell} - \frac{\dot \gamma^\mu }{\dot z}\frac{\partial z}{\partial \ell} \right]_{-s_\epsilon}^{+s_\epsilon} = \begin{pmatrix} \sinh\kappa\\ \cosh\kappa\\0\end{pmatrix},
\ee
and therefore
\be
\label{appAcfunc}
\widehat{{\cal F}}\left[\gamma(\ell)\right] = \ell ({\cal Q}_t[\gamma]\sinh\kappa + {\cal Q}_x[\gamma]\cosh\kappa).
\ee

\bibliography{biblio}{}

\end{document}